%% file: datalog-smt.tex
\documentclass[acmsmall]{acmart}
\setcopyright{none}

\usepackage{enumitem}
\usepackage{caption}
\usepackage{mymacros}
\usepackage{lang}
\usepackage{hyperref}
\usepackage{todonotes}
\usepackage{float}
\usepackage{subcaption}
\usepackage{multirow}
\usepackage{url}
\usepackage{siunitx}
\input{solidityenv.tex}

\AtBeginDocument{%
  }

\begin{document}

\newcommand{\oursystem}{\textsc{Desyan}}

\title{\oursystem{}: A Platform for Seamless Value-Flow and Symbolic Analysis}


\author{Panagiotis Diamantakis}
\email{pdiaman@di.uoa.gr}
\orcid{0009-0006-9730-5338}
\affiliation{%
  \institution{University of Athens}
  \city{Athens}
  \country{Greece}
}

\author{Thanassis Avgerinos}
\email{thanassis@di.uoa.gr}
\affiliation{%
  \institution{University of Athens}
  \city{Athens}
  \country{Greece}
}

\author{Yannis Smaragdakis}
\email{smaragd@di.uoa.gr}
\affiliation{%
  \institution{University of Athens}
  \city{Athens}
  \country{Greece}
}
\newcommand{\ethan}[1]{\todo{@Thanassis:~#1}}
\newcommand{\heavysmt}{Full SMT Solver\xspace}
\newcommand{\lightsmt}{Native Solver\xspace}



\keywords{TBD}
\ccsdesc{TBD}



\begin{abstract}
Over the past two decades, two different types of static analyses have
emerged as dominant paradigms both in academia and industry:
\emph{value-flow analysis} (e.g., data-flow analysis or points-to analysis) and
\emph{symbolic analysis} (e.g., symbolic execution). Despite their
individual successes in numerous application fields, the two approaches
have remained largely separate; an artifact of the simple reality that
there is no broadly adopted unifying platform for effortless and efficient integration
of symbolic techniques with high-performance data-flow reasoning.

To bridge this gap, we introduce \oursystem{}: a platform for writing program analyses
with seamless integration of value-flow and symbolic reasoning.
\oursystem{} expands a production-ready Datalog fixpoint engine
(\souffle{}) with full-fledged SMT solving invoking industry-leading SMT engines.
\oursystem{} provides constructs for automatically (and efficiently!)
handling typical patterns that come up in program analysis.
At the same
time, the integration is agnostic with respect to the solving technology,
and supports Datalog-native symbolic reasoning, via a bottom-up
algebraic reasoning module.

The result is an engine that allows blending different kinds of reasoning,
as needed for the underlying analysis. For value-flow analysis, the engine is the best-in-class Datalog
evaluator (often by a factor of over 20x in execution time); for applications that require full SMT (e.g., a concolic
execution engine or other symbolic evaluator that needs to solve arbitrarily
complex conditions), the engine is leveraging the leading SMT solvers;
for lightweight symbolic evaluation (e.g., solving simple conditionals
in the context of a path-sensitive analysis), the engine can use
Datalog-native symbolic reasoning, achieving large speedups (often of
over 2x) compared to eagerly appealing to an SMT solver.

\end{abstract}

\maketitle

\section{Introduction}

Static program analysis and verification (or just \emph{program analysis} in our context)
features an enormous variety of approaches, using different
formal tools, conceptual techniques, and implementation platforms. There is little similarity, in
concept and in realization, between \emph{pointer analysis}~\cite{PGL-014}, \emph{symbolic execution}~\cite{mattmight:King:1976:Symbolic,godefroid2005dart}, \emph{software model checking}~\cite{DBLP:journals/csur/JhalaM09}, and other sub-fields.
In the last two decades, in pointer analysis (and other
closely related \emph{flow analyses}, such as taint analysis) a dominant trend has emerged and driven
much progress in developing new approaches~\cite{aplas/WhaleyACL05,DBLP:conf/pldi/MadsenYL16,DataDriven2017,DBLP:conf/pldi/SzaboEB21,madsen13javascript,DataDriven2018,chord,oopsla/BravenboerS09,10.1145/2594291.2594327,Hybrid,Introspective,pldi/NaikAW06,aplas/LivshitsWL05}. The trend consists of specifying the analysis declaratively,
typically in a variant of the Datalog language. The declarative specification allows expressing
an analysis concisely, easily identifying its essence and variation points. At the same time,
high-performance Datalog engines (initially the proprietary LogicBlox~\cite{Aref:2015:DIL:2723372.2742796}, later the open-source \souffle{}~\cite{Jordan16}
and several research tools that it inspired~\cite{DBLP:conf/pldi/SzaboEB21,DBLP:conf/pldi/MadsenYL16}) have
ensured that the high-level specification becomes a highly-efficient implementation, often handily
out-performing hand-coded alternatives~\cite{oopsla/BravenboerS09}.

Given that flow analysis has found an elegant, unifying platform, it is tempting to add more
kinds of program analysis to the same platform. A single platform
enables \emph{blending} flow analysis with
different approaches in a unified analysis algorithm. Further, it helps
reap some of the same conciseness benefits in specifying analyses.

Symbolic execution is a prime candidate for blending with flow analysis. Both techniques
have found application to similar kinds of programs---often large, general-purpose imperative programs.
(This is in contrast to, for instance, software model checking, which has mostly found application
to distributed and concurrent programs, with emphasis on the complexity of non-deterministic interactions.)
Analysis techniques that use both value-flow reasoning and symbolic reasoning can be highly profitable,
with recent examples, such as Scuba~\cite{DBLP:conf/aplas/FengWDD15} (which leverages SMT solving in order to
summarize functions for points-to reasoning) and symbolic value-flow analysis~\cite{symvalic} (which has a value-flow
analysis structure but manipulates symbolic expressions).

In this work, we present \oursystem{}: an approach to blending symbolic reasoning and value-flow reasoning
\emph{seamlessly} in a single high-performance platform. \oursystem{} extends \souffle{}, the most-used,
highest-performance Datalog engine, with (a) functors that invoke an SMT solver, and (b) a native Datalog
symbolic reasoning library. (Each solver has its benefits in different kinds of symbolic reasoning and both
can be used interchangeably or mixed at will---e.g., one for small expressions, one for larger, both in the
same analysis.) The \oursystem{} extensions attempt to make the life of an analysis designer easier,
when interfacing with an SMT solver. They include rewrites to automatically introduce SMT idioms that
achieve high efficiency: expressing equality bindings via let-expressions and expressing ``bound variables''
with an inexpensive form of universal quantification (a pseudo-quantifier \forAllStar{}).

The result is a platform that can be used to express cutting-edge program analysis algorithms declaratively.
Flow analyses---the leading domain of declarative program analysis---are straightforward. If they are
already coded in \souffle{}, they run unmodified---no change in the language or evaluation semantics is
required---since \oursystem{} is just a \souffle{} extension using the standard \emph{functor} machinery.
Appealing to a symbolic solver is similarly seamless, with either the native Datalog solver or the
external-calling SMT functors. We demonstrate each of the approaches with a declarative symbolic execution
engine, an adaptation of symbolic value-flow (symvalic)~\cite{symvalic} analysis to \oursystem{}, and a
regular points-to analysis implementation.

In outline, the main contributions of the \oursystem{} work are:

\begin{bullets}
\item the seamless integration of SMT reasoning in a leading-edge, industrial-strength Datalog engine;
\item the automatic customization of SMT queries for common program analysis tasks;
\item the introduction of a Datalog-native ``lightweight'' symbolic reasoning engine that can serve as a
  predictable alternative to SMT for simple queries;
\item the study of the trade-offs of these approaches, for different styles of program analyses. In experimental terms, the integration with an industrial-strength engine yields speedups of over 7-26x
on average (for different configurations along the compiled/interpreted, single-/multi-threaded axes); the
Datalog-native symbolic reasoning engine achieves a 1.63x speedup over a large collection of inputs; the
use of SMT solving enables reasoning that is simply impossible through other means.
\end{bullets}

In total, these elements make for arguably the most powerful platform (as a combination of both
expressiveness and efficiency) for program analysis implementation and a model
of how to use it. The platform is fully open-source enabling unconstrained further development.

\section{Background}
\label{sec:background}

The declarative specification of program analysis algorithms has gained tremendous ground in the past two
decades~\cite{aplas/WhaleyACL05,DBLP:conf/pldi/MadsenYL16,DataDriven2017,DBLP:conf/pldi/SzaboEB21,madsen13javascript,DataDriven2018,chord,oopsla/BravenboerS09,10.1145/2594291.2594327,Hybrid,Introspective,pldi/NaikAW06,aplas/LivshitsWL05}.
The specification typically takes the form of Datalog rules, although the Datalog language is far from standardized
and its different variants offer distinct features and portability challenges~\cite{porting-doop}. Effectively,
the term ``Datalog'' has come to stand for any language with Prolog-like syntax and strictly declarative semantics
(i.e., reorderings of clauses inside rules and of rule evaluation do not affect the result, unlike Prolog). A
Datalog rule is simply a list of \emph{head} predicates instances that are derived every time the predicates in
the rule \emph{body} can be simultaneously instantiated. For instance, the rule:


\begin{rules}
\pred{HeadA}{x, y},\\
\pred{HeadB}{x, w, z} \rulearrow{} \\
\tab \pred{Body1}{x, y, z},\\
\tab \pred{Body2}{w, y}, \\
\tab \pred{Body3}{w, z}.\\
\end{rules}

\noindent will produce a tuple for \predname{HeadA} and another for \predname{HeadB} every time
there are values \args{w}, \args{x}, \args{y}, and \args{z} that simultaneously satisfy
\pred{Body1}{x, y, z}, \pred{Body2}{w, y}, and \pred{Body3}{w, z}.

Predicates in Datalog can be either EDB (Extensional Database) or IDB (Intensional Database). EDB relations provide the input facts of a program and IDB relations are derived from rules.
So, \predname{HeadA} and \predname{HeadB} have to be IDB, since they are defined by some rule, while \predname{Body1}, \predname{Body2}, \predname{Body3} can be either EDB or IDB.

The main appeal of declarative program analysis is the ease of expressing highly-complex recursive algorithms in compact form, while
achieving significant execution efficiency. Modern Datalog engines (e.g., \souffle{}) offer significant constant-factor improvements,
such as efficient computations over large volumes of data (through data organization that
admits excellent cache locality, much like in joins of database tables), and parallel execution (taking advantage of the declarative semantics
of the language, as well as the large-scale data operations over entire tables).
Even more importantly, modern Datalog engines support \emph{asymptotic} improvements of running time, via
automatic incrementalization of the computation. These transformations of the program analysis logic (to be rendered incremental) are
extremely hard to perform manually, therefore Datalog implementations are often much faster than equivalent hand-coded implementations
of the exact same algorithm. For instance, when the Doop framework was introduced, it boasted over 10x speedups compared to
pre-existing, manually optimized implementations that yielded the exact same analysis results~\cite{oopsla/BravenboerS09}.

\begin{figure}[tb!p]
\begin{tabular}{l}
\pred{VarPointsTo}{var, heap} \rulearrow{} \\
\tab \pred{Reachable}{meth}, \pred{Alloc}{var, heap, meth}. \\
\\

\pred{VarPointsTo}{to, heap} \rulearrow{} \\
\tab \pred{Move}{to, from}, \pred{VarPointsTo}{from, heap}. \\
\\

\pred{FldPointsTo}{baseH, fld, heap} \rulearrow{} \\
\tab \pred{Store}{base, fld, from}, \pred{VarPointsTo}{from, heap}, \pred{VarPointsTo}{base, baseH}. \\
\\

\pred{VarPointsTo}{to, heap} \rulearrow{} \\
\tab \pred{Load}{to, base, fld}, \pred{VarPointsTo}{base, baseH}, \pred{FldPointsTo}{baseH, fld, heap}. \\

\\
\pred{Reachable}{toMeth}, \pred{VarPointsTo}{this, heap}, \pred{CallGraph}{invo, toMeth} \rulearrow{} \\
\tab \pred{VCall}{base, sig, invo, inMeth}, \pred{Reachable}{inMeth}, \pred{VarPointsTo}{base, heap},\\
\tab \pred{HeapType}{heap, heapT}, \pred{MethodLookup}{heapT, sig, toMeth}, \pred{ThisVar}{toMeth, this}. \\
\\

\pred{InterProcAssign}{to, from} \rulearrow{} \\
\tab \pred{CallGraph}{invo, meth}, \pred{FormalArg}{meth, n, to}, \pred{ActualArg}{invo, n, from}. \\
\\

\pred{InterProcAssign}{to, from} \rulearrow{} \\
\tab \pred{CallGraph}{invo, meth}, \pred{FormalReturn}{meth, from}, \pred{ActualReturn}{invo, to}. \\
\\

\pred{VarPointsTo}{to, heap} \rulearrow{} \\
\tab \pred{InterProcAssign}{to, from}, \pred{VarPointsTo}{from, heap}. \\









\end{tabular}
\caption[]{Datalog rules for an Andersen-style points-to analysis and call-graph construction. The main computed
  relations are \pred{VarPointsTo}{var, heap} (a variable can point to a heap object);
  \pred{FldPointsTo}{baseH, fld, heap} (a ``base'' heap object can point to another via a field);
  \pred{CallGraph}{invo, meth} (an invocation instruction can dynamically call a certain method;
  \pred{Reachable}{meth} (a method is reachable and should be analyzed).
}
\label{fig:baserules-insens}
\end{figure}

For a glimpse of the compactness and expressiveness of a Datalog specification,
Figure~\ref{fig:baserules-insens} shows a standard Andersen-style~\cite{andersen:thesis}
points-to analysis as a sequence of 8 Datalog
rules. The analysis computes relation \predname{VarPointsTo} (a variable can point to a heap object) over a fairly powerful intermediate language (with virtual calls---\predname{VCall};
and heap load/store instructions---\predname{Load}/\predname{Store}).
The important element of this specification is that it is entirely realistic: this
is not a paper-only abstraction, but the actual implementation core of a high-performance
analysis. The analysis merely assumes that the input has been pre-processed into tables/relations,
that get produced straightforwardly. For instance, this input encodes program instructions,
such as \predname{Move}, \predname{VCall}, \predname{Load};
and standard symbol table and type system information, such as predicates \predname{FormalArg}, \predname{ThisVar}, \predname{HeapType}, and \predname{MethodLookup}.
Consider the first rule as an example. It states that if method \args{meth} is reachable and \args{meth} allocates to some variable \args{var} a heap object \args{heap}, then the rule derives that \args{var} points to \args{heap}. 
Note that \predname{Reachable} is an IDB predicate, as it is defined by another rule in the same program, while \predname{Alloc} is one of the input facts.

The 8 Datalog rules above are sufficient to
compute a powerful whole-program points-to analysis, including features such as
\emph{field sensitivity} and \emph{on-the-fly call-graph construction}.
Furthermore, the analysis can easily be extended, with fairly simple rule changes,
to admit significant precision enhancements, such as nearly any kind of
\emph{context sensitivity}~\cite{Sharir:Interprocedural,shivers:thesis,Milanova:2002:POS:566172.566174,1044835,10.1145/1926385.1926390}.
We refer the interested reader to a tutorial survey~\cite{PGL-014} for exploring pointer analysis in depth.

\section{\oursystem{} Illustration}
\label{sec:illustration}

For a simple (though partial) illustration of \oursystem{}, we show the
core of a \emph{symbolic execution} engine implementation. The rules
are standard Datalog rules up to the point of appealing to an external
SMT solver. Compared to a minimal formal specification of a symbolic execution
algorithm (e.g., \cite{5504796}) we add a realistic element:
computation on basic blocks (instead of single instructions), for input in
\emph{single static assignment} (SSA) form. These elements come from our
full-fledged \oursystem{} implementation of a real symbolic execution engine.
Importantly, we choose to show such elements in the illustration because
they enable us to raise important points about the declarativeness of the
\oursystem{} implementation and the order of evaluation.

\begin{figure}[tb!p]
 \begin{tabular}{l}
  \args{B} is a set of basic blocks  \\
  \args{E} is a set of binary expressions and constants\\
  \args{P} is a set of path conditions: lists of expressions \\
  \args{S} is a set of states: lists of basic blocks \\
  \args{V} is a set of variables \\
  \args{F} is a set of functions \\
  \args{O} is a set of (binary) operators \\
 \end{tabular}\\
\hspace{-3mm}
\begin{tabular}{l l}
\cline{1-2}
\pred{FunctionArg}{f:F, arg:V} & \# \args{arg} is a formal argument of function \args{f}\\
\pred{EntryBlock}{f:F, b:B} & \# \args{b} is an entry block for (public) function \args{f} \\
\pred{Assign}{b:B, v:V, c:E} & \# \args{v} is assigned to constant expression \args{c} in block \args{b}\\
\pred{PHI}{b:B, t:V, f:V}  & \# $\phi$ instruction in block \args{b} is ``\args{t} := $\phi(..., \args{f}, ...)$''\\
\pred{BinOperation}{b:B, o:O, l:V, r:V, s:V} & \# binary operation ``\args{s} := \args{l} \args{op} \args{r}'' exists in \args{b} \\
\pred{TrueEdge}{b:B, n:B, c:V} & \# \args{b} has branch over condition \args{c}; true edge to \args{n} \\
\pred{FalseEdge}{b:B, n:B, c:V} & \# \args{b} has branch over condition \args{c}; false edge to \args{n} \\
\cline{1-2}
\pred{Lookup}{s:S, v:V, e:E} & \# in state \args{s}, \args{v} has (symbolic expression) value \args{e}\\
\pred{BlockSetsVar}{s:S, b:B, v:V, e:E} & \# \args{b} reachable in state \args{s} sets \args{v} to \args{e} \\
\pred{Reachable}{s:S, p:P, b:B} & \# \args{b} is reachable in state \args{s} under path condition \args{p} \\
\end{tabular}
\caption[]{The domain, input relations, and computed relations for a
  symbolic execution engine.}
\label{fig:input-symeval}
\end{figure}

Figure~\ref{fig:input-symeval} establishes definitions of the domain
of computation and the input and output tables.
The symbolic execution computes three relations, \predname{Reachable}, \predname{BlockSetsVar}, and \predname{Lookup}.
The main output is \predname{Reachable}, which computes that a block can be reached under a sequence of
program conditions (``path conditions'' or ``path predicate'' in symbolic execution parlance), which are simultaneously satisfiable.
For this computation, the algorithm appeals to relations \predname{BlockSetsVar}, which captures the effects of a
single basic block over the state, and \predname{Lookup}, which captures the contents of a state, i.e., the mapping from
variables to values (i.e., symbolic expressions).

Predicate \predname{Lookup} (and the definition of a state, $s \in \args{S}$, as just a list of basic blocks) raises a subtle
point, which is good to keep in mind for declarative specifications:
States do not \emph{contain} their variable mappings. States merely get an \emph{identity}: when a state
is found to be reachable, it is generated as a unique id, by appending a block to the previous state. The state is then lazily populated with
variable mappings via evaluations of rules that add entries to the \predname{Lookup} relation.
That is, a state may be useful (e.g., to look up some variable) even when its \predname{Lookup}
table entries are still not complete: the analysis evaluation is free to execute the rules that populate \predname{Lookup}
in whatever order it wants, thanks to \souffle's semi-naïve fixpoint evaluation strategy. Symbolic execution in a declarative style is not really in-order ``execution'',
but instead as-parallel-as-possible evaluation
over a graph of data-flow dependencies.

\begin{figure}[tb!p]
\begin{tabular}{l}

\pred{BlockSetsVar}{stateBefore, block, var, value} \rulearrow{} \\
\tab \pred{Reachable}{stateBefore, \_, block}, \\
\tab \pred{Assign}{block, var, value}.\\
\\

\pred{BlockSetsVar}{stateBefore, block, to, expr} \rulearrow{} \\
\tab \pred{Reachable}{stateBefore, \_, block}, \\
\tab \pred{PHI}{block, to, from}, \\
\tab \pred{Lookup}{stateBefore, from, expr}.\\
\\

\pred{BlockSetsVar}{stateBefore, block, res, [op, leftExpr, rightExpr]} \rulearrow{} \\
\tab \pred{Reachable}{stateBefore, \_, block}, \\
\tab \pred{BinOperation}{block, op, left, right, res}, \\
\tab \pred{Lookup}{[block, stateBefore], left, leftExpr}, \\
\tab \pred{Lookup}{[block, stateBefore], right, rightExpr}.\\

\\\cline{1-1}\\

\pred{Reachable}{[entryBlock], nil, entryBlock} \rulearrow{} \\
\tab \pred{EntryBlock}{\_, entryBlock}. \\
\\

\pred{Reachable}{[block, stateBefore], [condExpr, pathCond], nextBlock} \rulearrow{} \\
\tab \pred{Reachable}{stateBefore, pathCond, jmpBlock}, \\
\tab \pred{TrueEdge}{jmpBlock, nextBlock, condVar}, \\
\tab \pred{Lookup}{[block, stateBefore], condVar, condExpr}, \\
\tab \smtResponse{\funcname{flatten}("AND", [condExpr, pathCond])} = ['sat'], \\
\tab |pathCond| < BOUND. \\
\\

\pred{Reachable}{[block, stateBefore], [newCondExpr, pathCond], nextBlock} \rulearrow{} \\
\tab \pred{Reachable}{stateBefore, pathCond, jmpBlock}, \\
\tab \pred{FalseEdge}{jmpBlock, nextBlock, condVar}, \\
\tab \pred{Lookup}{[block, stateBefore], condVar, condExpr}, \\
\tab newCondExpr = ["NOT", condExpr, nil], \\
\tab \smtResponse{\funcname{flatten}("AND", [newCondExpr, pathCond])} = ['sat'], \\
\tab |pathCond| < BOUND. \\

\\\cline{1-1}\\

\pred{Lookup}{[entryBlock], var, \funcname{fresh}()} \rulearrow{} \\
\tab \pred{EntryBlock}{fun, entryBlock}. \\
\tab \pred{FunctionArg}{fun, var}. \\
\\

\pred{Lookup}{[block, stateBefore], var, expr} \rulearrow{} \\
\tab \pred{Reachable}{stateBefore, \_, block}, \\
\tab \pred{BlockSetsVar}{stateBefore, block, var, expr}. \\
\\

\pred{Lookup}{[block, stateBefore], var, expr} \rulearrow{} \\
\tab \pred{Reachable}{stateBefore, \_, block}, \\
\tab !\pred{Assign}{block, var, \_}, !\pred{PHI}{block, var, \_}, !\pred{BinOperation}{block, \_, \_, \_, var},\\
\tab \pred{Lookup}{stateBefore, var, expr}.\\

\end{tabular}
\caption[]{Declarative rules for a symbolic execution engine in \oursystem{}, utilizing an SMT solver.}
\label{fig:rules-symeval}
\end{figure}

Figure~\ref{fig:rules-symeval} gives the full definition of the symbolic
execution algorithm. In terms of \emph{notation}: we use tuple
construction---[\args{element}, .. ]---as the basis of both lists and binary trees;
fresh symbolic variables are obtained via the \funcname{fresh}() constructor;
\funcname{flatten}(\args{conj},\args{list}) turns a list into a flat expression using conjunction operator \args{conj};
the size of a list is obtained via |\args{list}|;
``\_'' denotes a ``don't care'' value and ``!'' stands for negation.

This definition is very close to a mathematical specification of
forward symbolic execution~\cite{5504796}, but it is fully executable (modulo minor syntactic sugar) and
easily mixes in important realistic concerns, such as SSA input, block-level evaluation, and
stopping execution at condition length equal to a BOUND.

The rules are, for the most part, fairly simple. The rules for relation \predname{BlockSetsVar} express
straightforward evaluation of instruction semantics: for assignment of constants, for merging different
versions of a variable via a $\phi$ instruction of our Static Single-Assignment (SSA) input, and for binary operations.

The three rules for \predname{Lookup} associate a fresh symbolic variable with every public function argument,
establish a new binding for a variable in the ``after'' state of a block, if the block itself sets the variable,
and carry over the old binding if the block does not set the variable.

Finally, the three rules for \predname{Reachable} are the core of symbolic execution. They establish
transitive reachability of a basic block under a set of satisfiable conditions, by appealing to an
SMT solver (via the \oursystem{} key functor \smtResponse{}) on every conditional branch.

\myparagraph{Discussion.} The specification of the symbolic execution engine gives rise to several interesting observations:
\begin{bullets}
\item There is deep recursion between all three computed predicates. Each one of them
  is recursively defined over the others. This is the hallmark of declarative specifications
  in Datalog, much like in purely value-flow analyses (as in the many Datalog points-to analyses in
  the literature).
\item These recursive dependencies, together with the structure of the input program, are the
  determinants of evaluation order. The engine is free to evaluate the rules in whichever order
  it wants, which does not necessarily correspond to program execution order. As long as
  there are variable instantiations that satisfy the body of a rule, the inference in the head can be validly made.
\item Note in particular that the \predname{Lookup} of arguments of binary operations (third rule)
  is done on the \emph{after} state (\args{[block, stateBefore]}) of the block. This is valid because
  of the SSA input: each variable is assigned just once in the input form of the program. It is also
  necessary in the binary operation rule since the left or right operands may have been assigned in the same block.
  Thus, each variable can be looked up inside the basic block where it is assigned by checking the effects
  \emph{after} the block.
  (The exception to this ``after'' lookup is data-flow-merge, $\phi$, operators, which can occur at the
  beginning of a basic block and which may set variables that cyclically depend on others.)
\end{bullets}

As a result, we have a very compact executable declarative specification, amenable to high-performance
processing. The firing of a single rule can be processing many tens of program statements at once (as long
as inferring results for one statement does not depend on results of others that are currently also being
processed). All predicates are much like database tables, and the rule execution can be done in a (highly parallelizable
and locality-efficient) relational join of tables, in bulk. For instance, if the analyzed program has $N$ public functions (which correspond to
symbolic execution entry points), the symbolic execution will likely start processing all of them in the same evaluation step, and continue
as dictated by data-flow dependencies.

The core symbolic execution specification shown in Figure~\ref{fig:rules-symeval} is at the heart of a full-fledged
symbolic execution engine written in \oursystem{}. The full implementation (Section~\ref{sec:symeval-casestudy})
covers a complete, realistic
intermediate language with tens of instruction kinds, global memory of two kinds (transient and persistent),
and lots more.

\section{\oursystem{} Design}
\label{sec:design}

\oursystem{}'s top goal is to offer a platform for performing declarative-style program analysis
capable of performing symbolic reasoning without additional requirements.
\oursystem{} is built on top of \souffle{}, a high-performance, industrial-strength
Datalog implementation and requires no changes in the language or evaluation semantics.
Analyses written in \souffle{} can be dropped in to \oursystem{} and modified as desired to
incorporate symbolic reasoning.

\oursystem{} offers two separate components for symbolic reasoning (analysis designers can choose
one or the other, or mix them in the same analysis depending on the
tradeoffs available to them---see Section~\ref{sec:symvalic-eval} for an example):

\begin{bullets}
\item The \textbf{\heavysmt} integrates directly with SMT solvers and is best suited for complex symbolic reasoning workloads.
  Beyond the straightforward appeal to the solver, \oursystem{} incorporates several expressiveness
  and scalability elements of value to symbolic program analysis. These
  include let-expression generation and caching (for efficiency) as well as enabling custom operators
  like the pseudo-quantifier \forAllStar{} (for expressiveness).
  We detail the SMT solver integration in Section~\ref{sec:smt-integration}.
\item The \textbf{\lightsmt} is a pure-Datalog algebraic symbolic reasoner, for expressions up to
  bounded size. This solver results more predictably in high performance and works best when the
  queries are very simple, i.e., for lower-precision program analyses.
  We describe the Datalog reasoner in Section~\ref{sec:datalog-symbolic}.
\end{bullets}

\subsection{\heavysmt with Functors}
\label{sec:smt-integration}

The main element of \oursystem{} is the interfacing between \souffle{} Datalog and SMT solvers.
Mechanism-wise, this is achieved via \souffle{} ``functors''~\cite{functors}: strongly-typed functions, implemented in C/C++.
In order for \souffle{} to communicate with an SMT solver, the functors need to transform queries into a specific
format, use the solver's API to send the query, and receive and decode the response. The illustration of Section~\ref{sec:illustration}
provides a simplified view of the main functor, \smtResponse{}, but elides many important elements, discussed next.

\myparagraph{\smtResponse{}.} This is the main functor for SMT solver invocations inside a Datalog rule.
The functor takes as input a query, written in the SMTLIB format~\cite{barrett2010smt}.
Functors need to follow strict pure functional semantics, in order to be well-integrateable in the
evaluation of a declarative program: invoking a functor multiple times should return the same
result and the fact of the invocation (or not) should not be observable inside the Datalog rule evaluation.

The query response includes the ``sat''/''unsat'' answer and, in the former case, the model, i.e., variable
assignments, establishes satisfiability. This is represented as a Datalog tuple and can be used as
a first-class citizen inside Datalog rules.

\myparagraph{\printToSmt{}.} The second functor employed in \oursystem{} is \printToSmt{}.
Since \smtResponse{} expects a query written in SMTLIB format, \printToSmt{} is used to handle the translation from the
Datalog representation of logic formulas into valid formulas in SMTLIB form. In our
illustration of Section~\ref{sec:illustration}, the complexity of this functor was unnecessary, so it is hidden
behind a simple syntactic sugar operator, \funcname{flatten}. In reality, the functor does more than a straightforward
flattening of the input---e.g., ``custom operators'' and ``automatic let-bindings'', discussed next.

\oursystem{}'s \heavysmt comes with several convenience features for symbolic program analysis:

\begin{bullets}
\item{\textbf{Standard SMT Logics.}} The \heavysmt{} component utilizes Z3~\cite{demoura2008z3} in the
  backend, which means that all Z3-supported logics are available for symbolic program analysis
  in \oursystem{}. Most of our experiments (Section~\ref{sec:experiments}) operate on constraints
  using standard BitVector (BV) and Quantifier Free BV (QF\_BV) logics, but our team has had
  success using \oursystem{} with more expressive logics as well.

\item{\textbf{Custom Operators and the \forAllStar{} Quantifier.}} \oursystem{}'s design enables
  analysis designers to introduce their own custom operators, which can be used in the SMT queries.
  This is a powerful feature that can be used to define complex operations that are not directly
  supported natively by the SMT solver. Custom operators can both improve the expressiveness of the
  analysis and also help optimizing solver performance.

  As a use-case, we used \oursystem{}'s custom operators to define a pseudo-quantifier, \forAllStar{}, which is a
  lightweight-\textbf{forall} ($\forall$) quantifier: an extra argument
  specifies the variables that should be bound using it.
  The \forAllStar{} quantifier gives a heuristic but effective solution to a common problem that we have identified in program analysis:
  the issue of ``bound'' variables.
  Bound variables are a concept that arises for any analysis that does not involve just concrete values. The meaning
  of a bound variable is ``an unknown quantity, yet one that should not be considered controllable for the
  analysis task being performed''. For instance, consider the following code snippet:
  \begin{verbatim}
        bool deleteAdmin(int userid) {
          if (userid == ADMIN) {
            // sensitive action
          }
        }
  \end{verbatim}

  \noindent The program variable \sv{userid} is used to designate the identity
  of a logged-in user. The \sv{userid} is checked against a special value \sv{ADMIN}, which designates a
  trusted party that can perform authorized actions. When performing an analysis for vulnerability detection,
  the SMT query should not treat \sv{userid} as a variable that it can freely set, or the program would appear
  trivially vulnerable: the admin can indeed perform sensitive actions. Instead, the true logical query one often wants to perform
  is of the form ``for all values of \sv{userid}, do values of free variables exist such that...?''
  This is certainly possible with \oursystem{}, however SMT solvers
  are notoriously hard-pressed to deal with ``for all''-style queries: universal quantification gets (typically)
  translated into validity, instead of satisfiability, queries, which end up easily unscalable.
  Since the exact value of \sv{userid} is hardly relevant
  (provided that it not be equal to the \sv{ADMIN} constant), one can do much better in an automatic way
  by utilizing \oursystem{}'s custom operators.

  The \forAllStar{} custom quantifier prevents the solver from assigning a specific value to the bound variable, in its effort
  to satisfy the formula.
  Instead, \forAllStar{} binds the variable to a randomly chosen value from a predefined pool of ``magic constants''.
  These are special values of large size (256-bit vectors) with partly human-recognizable and partly irregular
  numeric/bit patterns, so they are virtually guaranteed
  to not be derivable through regular arithmetic or program operations that involve other unrelated values, including
  each other.

  This approach leads to two possible cases. In case of unsatisfiability, we can be certain that the formula would not hold
  even if the standard \textbf{forall} were used. However, in case of satisfiability, we have not definitively
  proven the formula would be equally satisfiable with a regular \textbf{forall}: it is conceivable, although
  unlikely, that the
  magic constant somehow has contributed to finding a satisfying assignment, which would not have been possible
  for some other value of the \forAllStar{}-quantified variable.

  In practice, we have found the \forAllStar{} quantifier to be
  highly effective at getting desirable solutions efficiently, without ever producing models that artificially match the magic constants.
  (However, should a human inspect the model, the magic constants and values derived from them should be easy to
  identify, due to their recognizable bit patterns. Thus one can inspect and be certain of the model's independence
  from the bound variables.)

  In the following examples (Listings \ref{lst:forall-example}, \ref{lst:forallstar-example}) we show a simple query quantified with \textbf{forall} and \forAllStar{}, respectively.
  In the first case, the solver returns ``\sv{unsat}'' in $\sim$3 seconds, while in the latter case in $\sim$0.03 seconds

  \begin{lstlisting}[caption={\textbf{forall} example written in SMTLIB format}, label={lst:forall-example}]
(assert
  (forall ((x  (_ BitVec 256)))
    (=  #x01
        (bvsub
          (bvurem (bvmul x  x)
                  #xff...ff)
          #x00))))
\end{lstlisting}

\begin{lstlisting}[caption={\forAllStar{} example written in SMTLIB format}, label={lst:forallstar-example}]
(declare-const x (_ BitVec 256))
(assert
    (=  #x01
        (bvsub
          (bvurem (bvmul x  x)
                  #xff...ff)
          #x00)))
(assert (= x #x1123456789abcdef0123456789abcdef...0123456789abcdef))
\end{lstlisting}


\item{\textbf{Query Caching.}}
  Since an SMT solver can be non-deterministic (i.e., return a different solution for the same query), the
  first requirement for \oursystem{} functors is to have an ``oracle'' form in their implementation:
  queries and their responses are cached, and if a query has been submitted previously, the cache
  is consulted so that the earlier results gets returned. This feature also helps with evaluation efficiency.

\item{\textbf{Automatic Let-Bindings}}
  \oursystem{} provides an option for automatically introducing let-bindings in formulas generated by a
  symbolic executor for clauses that involve equality constraints. Consider a symbolic execution engine,
  as in Section~\ref{sec:illustration}. The current state associates variables with symbolic expressions
  (the current ``value'' for the variable). Imagine a state that has a path predicate of the form
  $ \sv{x1} = \sv{fresh} \land \sv{x2} = \sv{x1} + 42 \land \sv{x1} + \sv{x2} = 16 $, with any satisfiable assignment
  to \sv{fresh} indicating the presence of a bug (the analysis goal). By activating the let-bindings option and passing
  let-bound variables to it (\sv{x1}, \sv{x2} in this case), the solver will automatically turn the query above into
  $ let\: \sv{x1} = \sv{fresh} \: in \: let \: \sv{x2} = \sv{x1} + 42 \: in \: \sv{x1} + \sv{x2} = 16 $.
  Let-bindings help reduce the solver's search space (only $\sv{fresh}$ is free in the example above) and
  typically lead to better performance.
  This feature is optional, since analysis writers may want to have full control of the query and how
  let-bindings are introduced in the final formula.
\end{bullets}

\myparagraph{Example Translation.}

In Figure~\ref{fig:input-symeval} we defined the domain of formula expressions as:
``\args{E} is a set of binary expressions and constants'' which translates to the following
\souffle{} Datalog type definition:
\begin{lstlisting}[caption={\souffle{} type of logical formulas}]
.type PrimitiveExpr = symbol
.type Operator = symbol
.type Base = PrimitiveExpr | Operator
.type Expr = [
         base: Base,
         left: Expr,
         right: Expr
     ]
\end{lstlisting}
Primitive expressions can be either constants (e.g., \sv{"0xf"}) or variables (e.g., \sv{"var"}).

As operators, we consider a rich set of numeric operators from a real-world intermediate representation.
These, in many cases, translate directly to SMTLIB functions (e.g., \sv{"ADD"} into \sv{bvadd}).
In cases when the operator is not supported, we could declare custom functions
and add that declaration in the start of the query. Instead, to achieve better performance, \oursystem{} inlines
the definition of operators. For instance, the following code (inside the functor implementation) generates
an \sv{ite} (``if-then-else'') clause that implements a 256-bit-result \sv{isZero} check.

\begin{lstlisting}[caption={Example of inlined user defined bitvector function}]
#define SMTLIB_TRUE_VAL
  "#x0000000000000000000000000000000000000000000000000000000000000001"
#define SMTLIB_FALSE_VAL
  "#x0000000000000000000000000000000000000000000000000000000000000000"

string inline_user_defined_function(string op, string lexpr, 
                                    string rexpr) {
 if(op == "isZero") {
     return "(ite (= "+ lexpr +" "+ SMTLIB_FALSE_VAL +")"
             " " + SMTLIB_TRUE_VAL
             " " + SMTLIB_FALSE_VAL
             ")";
 }
 ...
}
\end{lstlisting}

As a complete example, a program in \oursystem{} solving the logical formula
$ let\: \sv{x1} = \sv{fresh} \: in \: let \: \sv{x2} = \sv{x1} \ll 1 \: in \: \sv{x1} + \sv{x2} = 3 $ (where $\ll$ is the Shift Left
(SHL) operator) given as a query to a  \souffle{}
will look as follows:
\begin{lstlisting}[caption={Simple rule introducing let bindings},label={lst:Rule-lets}]
Query(response) :-
  constr =
    ["EQ", ["ADD", ["x1", nil, nil], ["x2", nil, nil]],
           ["0x03", nil, nil]],
  lets =
    [["x2", ["SHL", ["x1", nil, nil], ["0x01", nil, nil]]],
    [["x1", ["fresh", nil, nil]], nil]],
  response = @smt_response_with_model(@print_to_smt(constr, nil, lets)).
\end{lstlisting}

The corresponding SMTLIB formula produced by \printToSmt{} is:

\begin{lstlisting}[caption={SMTLIB-formatted query of Listing \ref{lst:Rule-lets}} with numeric constants condensed from 256 bits to 8 for readability.]
(declare-const fresh (_ BitVec 256))
  (assert
    (let ((x1 fresh ))
    (let ((x2 (bvshl x1 #01)))
      (= #01
         (ite
          (= (bvadd x1  x2 )
             #03)
          #01
          #00
      )))))
\end{lstlisting}

while the result from the \smtResponse{} would be:

\begin{lstlisting}[caption={The result calculated by rule in Listing \ref{lst:Rule-lets}}]
[sat, [[fresh, 0x1], nil]]
\end{lstlisting}

\myparagraph{Implementation.}
The goal of \oursystem{} is to offer a seamless, industrial-strength integration of Datalog
rules and symbolic reasoning. The choice of implementation platform as well as the features supported
by the functors (e.g., caching, let-bindings, bound variables) aim precisely at enabling optimal, easy-to-write,
state-of-the-art program analysis implementations.

We use Z3~\cite{demoura2008z3} as the default SMT solver for \oursystem{}. \oursystem{} functors are fully integrated
in the \souffle{} Datalog implementation, following best practices both on the \souffle{} end and on the Z3 end.
For instance, when a
functor that has symbols as arguments is called, \souffle{} passes their ordinal numbers instead.
Therefore, in the functor implementation in C/C++, \oursystem{} makes use of the \souffle{} class \sv{SymbolTable},
and its functions \sv{encode/decode} that translate the ordinal into the string that it represents.
Similarly, \oursystem{} uses class \sv{RecordTable} to manage records through the functions \sv{pack/unpack}, and
expose as native Datalog tuples the results of an SMT query.



\subsection{\lightsmt in Datalog}
\label{sec:datalog-symbolic}

The second solver available in \oursystem{} is a symbolic solver implemented purely declaratively, in Datalog.
This cannot replace the integration with SMT solvers but can complement it, since it is a fundamentally
different mechanism. Conceptually, SMT solvers are ``top-down'' search algorithms: they seek \emph{one} solution in an
exponentially large (and, thus, not exhaustively searchable) search space. In contrast, Datalog
computations are ``bottom-up'' algorithms: they produce \emph{all} valid inferences that follow from a set of
rules and input facts. Datalog computations are typically bound to be polynomial---in fact, the base Datalog
language is guaranteed to only express polynomial algorithms, although realistic implementations have several extensions
that render them Turing-complete. As a result of this fundamental property, Datalog computations are highly efficient
when most (or all) of their results are needed. The computation can be performed in bulk, by joining large tables
in a highly-parallel fashion, as already discussed.

Accordingly, the pure-Datalog solver in \oursystem{} takes advantage of the efficient bulk processing in the language in order to produce
all valid inferences from a set of algebraic axioms.  The main application is to produce simplifications/normalizations of a formula
of bounded size (e.g., up to 10 nodes, in tree form). This is a valuable approach for simple, local symbolic reasoning, with excellent
predictability. For instance, to see if a program branch condition is satisfiable with current symbolic and concrete values
\emph{independently} of other branches in the program, the pure-Datalog solver is the appropriate mechanism to appeal to.
The same query can be answered by an SMT solver query, but the cost can be large and hard to predict (especially in the presence
of large bit-vector operations). In contrast, for large formulas (e.g., an entire path condition of a symbolic execution),
the SMT solver is the only realistic option---the Datalog solver cannot scale to even a fraction of the expression size.
(We discuss with concrete numbers in Section~\ref{sec:experiments}.)

Concretely, the pure-Datalog solver of \oursystem{} accepts as input from a program analysis an initial set
of constants and symbolic expressions, together with a designation of symbolic variables as free or bound.
The solver then defines a universe of evaluation from these initial values by closing
over the following sets of reasoning rules:
\begin{bullets}
\item For every expression, all its subexpressions are in the universe of evaluation.
\item For every expression, if an algebraic simplification rule produces a smaller expression,
  the new expression is also in the universe. For expressions with syntactic trees of the same size,
  ``smaller'' is an arbitrary ordering relation.
\item For every constant, its logical complement and equality with another constant are
  expressions (which will be constant-folded).
\item Constant folding is performed to yield new constants, if both arguments of an
  expression (already in the universe of evaluation) are constants.
\end{bullets}

Based on the above, the pure-Datalog symbolic reasoner is guaranteed to terminate: the universe of
expressions is bounded and as a result it also bounds the amount of constant folding that can be performed,
since constant folding can only specialize an existing expression.

The algebraic axioms of the solver are rich but largely standard:
\begin{bullets}
\item Operators are classified as associative, commutative, left/right distributive, idempotent, canceling (when
  applying the operator to the same element twice yields a constant), mutually exclusive (when one binary operator
  implies the negation of another). Inverse operators are additionally matched in left/right pairs.
\item Special values (e.g., identity and zero elements) for every operator are defined.
\item All the standard algebraic rewrite axioms for the above categorization of arithmetic and logical operators
  are known to the solver and are fully applied. Extra, ad hoc algebraic axioms are available in some cases
  (e.g., for multiplication/division by a power of 2).
\item The solver has solution strategies for linear equalities and inequalities, for both arithmetic and logical
  operators.
\end{bullets}

For illustration, the potential rewrite of an associative operator is expressed (modulo minor syntactic sugar)
as a simple rule:

\begin{rules}
\pred{BaseEquals}{expr, [op, leftleft, [op, leftright, right]]} \rulearrow{} \\
\tab \pred{IsExpression}{expr}, \\
\tab \args{expr} = \args{[op, [op, leftleft, leftright], right]} \\
\tab \pred{AssociativeOperator}{op}. \\
\end{rules}

(The \predname{BaseEquals} relation is the base relation that will get transitively closed
to produce the final \predname{Equals} result of the symbolic solver.)


As a slightly more complex example, the solver defines which operator can provide a solution for linear expressions
over another operator, via predicate \predname{LinearSolutionOperators}---e.g., the declaration

\begin{rules}
\pred{LinearSolutionOperators}{"ADD", "SUB"}.\\
\end{rules}

\noindent means that ``\sv{x + [y] = [z]}'' (where only \sv{x} is a free variable and \sv{[y]}, \sv{[z]} are arbitrary
expressions without \sv{x}) can be solved by the value ``\sv{[z] - [y]}'' for \sv{x}.

The solver also defines less intuitive solution operator combinations, such as:

\begin{rules}
\pred{LinearSolutionOperators}{"MOD", "ADD"}.\\
\pred{LinearSolutionOperators}{"OR", "OR"}.\\
\end{rules}

The core rule of the rewrite strategy for linear solutions (applied after expressions
are normalized to move free variables to the left) is then simply:

\begin{rules}
\pred{ValueForFreeVariable}{leftleft, [opInverse, right, leftright]} \rulearrow{} \\
\tab \pred{IsExpression}{["EQ", [op, leftleft, leftright], right]}, \\
\tab \pred{LinearSolutionOperators}{op, opInverse}, \\
\tab \pred{IsFreeVar}{leftleft}, \\
\tab \pred{NoFreeVarExpression}{leftright}, \\
\tab \pred{NoFreeVarExpression}{right}. \\
\end{rules}


The above aspect of the native Datalog reasoner is limited top-down (goal-directed) search for
solutions, disguised behind a bottom-up reasoning technique that produces all valid
formulas. The key is that an equality in the original program together with a free variable trigger
(i.e., perform in a goal-directed way) a limited rewrite that ``solves'' a constraint for the free
variable. Thus, one can think of the native Datalog solver as a bottom-up algebraic rewrite system
combined with a very limited capacity to ``solve'' linear equalities, just as a top-down solver
would do.

\myparagraph{Interaction with analysis.}
It is worth considering how a program analysis algorithm can use the lightweight pure-Datalog symbolic solver.
The solver itself is guaranteed to terminate (for a single instantiation of its initial constants
and symbolic expressions). However, the solver creates new expressions (through algebraic rewrites), as well as suggests values for
free variables. Both of these outputs can give rise to even more expressions when propagated through the program.
Since the program will likely have cyclic data-flow dependencies, its analysis cannot simply be
mutually recursive with the Datalog solver without any further constraints: if fresh expressions from the solver
were to always propagate, get enriched through program operations treated symbolically, and propagate back to
the folder, the process would not terminate. Fortunately, this cannot happen by accident: the solver performs
expression normalization (producing the smallest equivalent expression), which is a non-monotonic operation and, thus,
requires its execution to be ``stratified'' relative to the rules that produce its inputs:
Datalog does not permit recursion from the solver's outputs to its inputs.

However, this causes the converse problem: an analysis cannot employ the solver, receive solutions, and feed
them back to the solver for more solutions. To address the latter, \oursystem{}'s pure-Datalog solver's is
implemented as a \souffle{} Datalog ``component''.\footnote{\url{https://souffle-lang.github.io/components}} A
\souffle{} component can be instantiated any number of times (possibly with different parameterizations), producing distinct instances
of isomorphic rules. This allows the solver to be
employed any bounded number of times, with each instantiation receiving results after the analysis
processes the output of the previous one.

\section{Case Studies and Evaluation}
\label{sec:experiments}

We next present example program analyses leveraging \oursystem{}. Although the discussion is experiments-oriented,
it contains more qualitative remarks than a straightforward experimental evaluation.
\oursystem{} enables a combination of features (mature Datalog evaluator + SMT + specialized
solver) not encountered before. Omitting any of these features may result in an analysis being
infeasible or can demonstrate ``unfair'' benefit.

\subsection{Points-To Analysis}
\label{sec:experiments-points-to}

The first design element of \oursystem{} is that it builds over an industrial-strength, mature
Datalog engine, \souffle{}. This is in contrast to other approaches of research interest,
such as the Formulog system~\cite{DBLP:journals/pacmpl/Bembenek0C20,DBLP:conf/datalog/Bembenek0C22,formulog24}.
Formulog has offered a research platform for blending together functional programming,
Datalog rules, and SMT queries.


This results in a substantially different approach from that of \oursystem{}, which emphasizes strong, seamless support
for expressing the full program analysis declaratively, as Datalog rules, without the possibility
of easily escaping into a conventional form of analysis specification (via the use of functional
programming). Formulog still
represents an interesting direction since it raises important questions regarding the
match or mismatch between these paradigms---for instance, recent work on ``making Formulog fast''~\cite{formulog24}
has led to the question of whether it is fruitful to occasionally abandon one of the main implementation
tenets in efficient Datalog engines, semi-naive evaluation.

However, Formulog is currently not a realistic platform for building state-of-the-art program
analyses. The authors of Formulog recently ``\emph{provid[ed] strong evidence that Formulog can be the basis of a
realistic platform \textbf{for SMT-based static analysis}}''~\cite{formulog24}. Yet the language
is not a realistic platform for value-flow analyses---the domain where declarative program analysis
has excelled. Porting a Datalog-based analysis (e.g., that of Section~\ref{sec:background})
to Formulog results in over 10x slowdowns, even for relatively modest inputs.

To evaluate this aspect, we implemented a complete but compact points-to analysis in both \oursystem{} and Formulog.
The analysis is a straightforward adaptation of the Doop framework's~\cite{oopsla/BravenboerS09} ``micro'' analysis: for the
\oursystem{} version the analysis runs unmodified;
for Formulog, the analysis is a direct transcription. We start the analysis from
methods with the usual Java \sv{main} signature.
We conduct the analysis on a machine with two Intel(R) Xeon(R) Gold 6136 CPU @ 3.00GHz (each with 12 cores x 2 hardware threads)
and 640GB of RAM, and evaluate over the DaCapo benchmarks~\cite{dacapo:paper}.

In order to give a comprehensive overview of the performance of the two systems, we include four different configurations:
interpreted single-threaded, interpreted multi-threaded, compiled single-threaded and compiled multi-threaded.


\begin{figure*}
  \centering
  \begin{subfigure}[b]{0.8\textwidth}
      \centering
      \includegraphics[width=\textwidth]{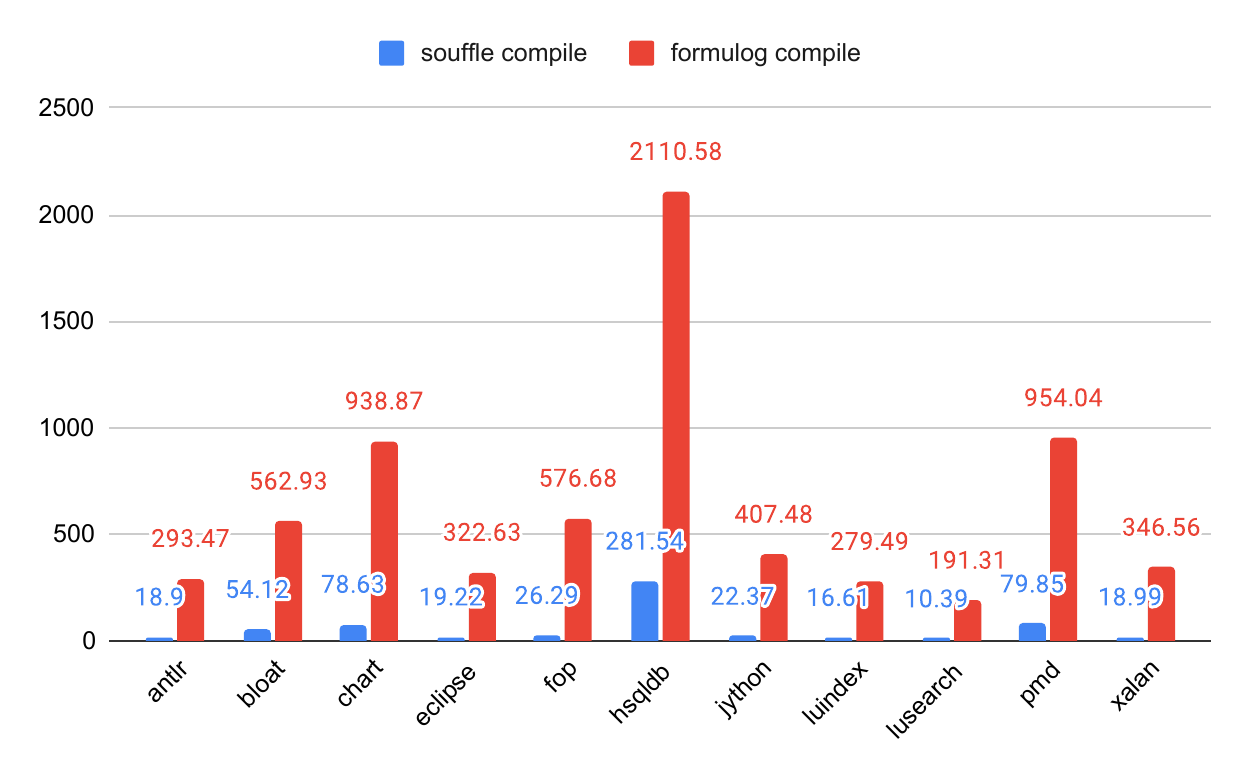}
  \end{subfigure}
  \caption[]
          {Execution time of Point-To analysis (in sec), comparison of \souffle{} vs. Formulog.
            Compiled sequential version.}
  \label{fig:times-per-benchmark-per-config-seq-compiled}
\end{figure*}

Figure~\ref{fig:times-per-benchmark-per-config-seq-compiled} shows the average analysis times (for a single run, but we have neither
observed large variation, nor are the qualitative results affected by small changes)  for the compiled single-threaded configuration.

As shown in ~\ref{appendix:experimental-results}, the average speedup across all benchmarks ranges from $\times 7$ to $\times 26$.

To reiterate, the above experiment is not reflecting an outright contribution of \oursystem{}:
the performance is transparently inherited from \souffle{}. However, the experiment
reflects \emph{why} \oursystem{} is a new, important platform compared to other efforts:
there is no past analogue to \oursystem{}'s ability to handle both high-performance declarative
analyses and symbolic reasoning.

\subsection{Lightweight Symbolic Reasoning: Symbolic Value-Flow Analysis}
\label{sec:symvalic-eval}

Symbolic value-flow (\emph{symvalic}) analysis~\cite{symvalic} is a recent blend of value-flow
and symbolic reasoning, applied with commercial success to the domain of Ethereum VM smart contracts~\cite{symvalic-sbc23}.
The analysis employs simple symbolic reasoning entirely inside Datalog. This
has been a major inspiration behind the native Datalog symbolic solver
of the \oursystem{} work (Section~\ref{sec:datalog-symbolic}). Effectively, \oursystem{} proposes
that one can generalize native Datalog symbolic reasoning, abstract it away from analysis
specifics, and offer it as a platform for efficient lightweight symbolic reasoning.

Symvalic analysis performs symbolic reasoning but over fairly simple expressions. Recall
that a standard symbolic execution engine (cf. Section~\ref{sec:illustration}) maintains a full
symbolic \emph{path condition} containing the conjunction of all conditional expressions that need
to be satisfied in order to reach a certain program point. In contrast, symvalic analysis examines the
individual conditions independently. Once a condition is found to be satisfiable, the analysis
just records the combination of (possibly symbolic) values that satisfy the condition and proceeds
to the next. The combinations are grouped into value sets based on \emph{dependencies}: value bindings for
function arguments and most-recent loaded values from global memory.

To see an example of how symvalic analysis loses precision (compared
to concrete or symbolic execution of the program), consider the code fragment below:

\begin{javanonumbercode}
function deposit(uint 
  uint actualFees = min(fees, allowed[recipient]);
  if (actualFees > 200)
    if (fees < 100)
      ...
\end{javanonumbercode}

The two conditions are not simultaneous satisfiable: \sv{actualFees} is at most \sv{fees},
therefore if \sv{actualFees > 200}, then it cannot be the case that \sv{fees < 100}.
A regular execution (symbolic, concrete-symbolic, or just concrete) would detect this
infeasibility.

For symvalic analysis, however, precision is only kept up to dependencies.
The assumption in the example (reflected via boldface in the listing) is that the analysis has picked
parameters \sv{amount} and \sv{recipient} as the basis of dependencies for analyzing function \sv{deposit}.
(Picking the basis of dependencies is heuristic, but the number of elements in the basis is certainly
fixed to a small
constant, typically 2 or 3. Therefore, many elements of the current execution state will not be
reflected in dependencies.)
This means that the function is analyzed separately for all combinations of values that the analysis has
derived for \sv{amount} and \sv{recipient}. Values of \sv{fees}, however, are only treated as a
set. Therefore, if the analysis was to have inferred as possible values for \sv{fees} the
set $\{50,300\}$ and the value of \sv{allowed[recipient]} was $500$, then in the last statement
of the example the possible value set for \sv{actualFees} would be $\{300\}$, but the value set
for \sv{fees} would remain $\{50,300\}$, making the analysis consider the last condition satisfiable:
the dependency between \sv{actualFees} and \sv{fees} fails to be captured. (If \sv{fees} had been
part of the dependencies for the analysis of the function, the values inferred for
\sv{actualFees} would have been of the form $\{300 |$ if \sv{fees} $= 300\}$.)

Symvalic analysis makes this precision tradeoff in order to gain scalability and completeness:
the analysis covers a lot more program statements, by not having to fully solve all conditions
needed to reach them. The specifics of whether this precision-completeness-scalability tradeoff
pays off are orthogonal to our discussion, but what is pertinent is the simplicity of
symbolic conditions that arise. Since every conditional in the analyzed program is treated
separately, symbolic solving is much lighter-weight than in a conventional
symbolic execution engine. The form of queries remains the same, with a need to
provide a model (i.e., find values for free variables that lead to condition satisfiability)
but the complexity of expressions is a lot lower.

We adapted the full symvalic analysis to use \oursystem{}, both by invoking the SMT
solver and by using the native Datalog solver. The choice is determined via analysis-time
parameters and the same analysis can use both solvers. Specifically, the main parameter
determines at which symbolic expression size to switch from the native solver to the
SMT solver. (Therefore, the analysis can choose to use exclusively one of the two solvers
by setting the parameter to zero or to a high number.)

The integration with the SMT solver makes full use of the \forAllStar{} quantifier,
for bound variables, and exercises both satisfiability and validity queries. (Validity queries
are used for expression equivalence checks and have the expected form: they invoke the \smtResponse{}
functor with a negated query and check for  ``\sv{unsat}'' answer.)

Somewhat surprisingly, we did not find a profitable setting for using the SMT solver with
symvalic analysis. SMT solving was uniformly slower than native solving. In expressions
under the size limit of the native solver, the use of the SMT solver yields equivalent
results for higher cost. If applied to larger expressions, the analysis gained very little benefit
for a significant cost. This observation
is certainly explainable, given that the analysis is designed to only operate on simple
symbolic conditions.

To quantify the benefit of the native solver we consider a standard benchmark set for
symvalic analysis. This consists of 189  ``substantial'' contracts, as selected by
the authors of the Gigahorse decompiler~\cite{189-contracts}.

Figure~\ref{fig:symvalic-all} shows the results in table form. The native solver is significantly
faster and, more importantly, succeeds in analyzing all contracts in the input set.
For these contracts the average number of queries to the solver is 1,115 (max: 13,487).
In this setting (many small queries) the lightweight native solver showcases its value.
The SMT solver is in most cases entirely adequate but its unpredictability is apparent
in the number of contracts that time out.

\begin{figure}
\begin{center}
\begin{tabular}{|l|c|c|}
  \hline
     & \lightsmt & \heavysmt \\
  \hline
  failed to analyze & 0 & 7 \\
  average analysis time & 129.6sec & 218.1sec \\
  \hline
\end{tabular}
\caption[]{Performance of the \lightsmt vs. \heavysmt for symvalic analysis of 189 smart
contracts.}
  \label{fig:symvalic-all}
\end{center}
\end{figure}

It is this unpredictability that motivates having a native Datalog solver. As can be seen
from the above numbers, the \emph{average} cost of using the SMT solver for small
symbolic expressions is low: the extra time taken (88.5sec on average per contract)
means the average overhead per query (1,115 queries for the average contract) is under 80ms.
However, a small number of queries can have significantly higher cost.
(This is a fact commonly acknowledged in the literature, which we are also documenting in our
setting for completeness.)

For instance, the query below is one arising as part of the reasoning in one of the contract analyses.
\begin{lstlisting}[caption={Example query that has significant cost for the SMT solver. The
      numeric constants are condensed for presentation: every hexadecimal value is 256-bit long,
      i.e., 62 leading zeros have been truncated from all shown constants.}]
(declare-const <<owner-unique-value>> (_ BitVec 256))
(declare-const adota (_ BitVec 256))
(declare-const xspacexquotequote (_ BitVec 256))
(declare-const yspacey (_ BitVec 256))
(declare-const zcolonz (_ BitVec 256))
(assert
 (= #x01
  (ite (= #x00
          (bvsub (bvudiv #x01 <<owner-unique-value>>) #x00))
   #x01 #x00)))
(check-sat)
\end{lstlisting}

The query takes $\sim$5sec on the Z3 solver, largely due to reasoning about
the bit vector division operation. Experienced users of SMT solvers can often avoid such costly
queries, by changing representations (e.g., perhaps mathematical integers, instead of 256-bit ones,
are sufficient for the performed analysis) or reformulating the query (e.g., perhaps a division
constraint has an easy solution if formulated as multiplication). However, no \emph{general} technique
exists to always predictably guarantee an efficient rewrite. Furthermore, performance problems need to
be identified per-case, which is not always possible when one builds a general analysis algorithm,
over very diverse input programs.

\subsection{Symbolic Execution}
\label{sec:symeval-casestudy}

Section~\ref{sec:illustration} discussed the core of a symbolic
execution engine in \oursystem{}. We have built a full symbolic
executor for the three-address code intermediate language of the
Gigahorse decompiler~\cite{gigahorse,elipmoc} for Ethereum VM smart
contracts---the same input as that of the symvalic analysis. This is a rich
intermediate language with many arithmetic instructions over 256-bit unsigned
numbers. The size of the full implementation is a mere \textasciitilde{}60 \oursystem{} rules or
\textasciitilde{}500 lines of code.

The symbolic execution engine is only possible because of the SMT support in \oursystem{}.
The conditions that arise are well beyond the capabilities of the native Datalog symbolic solver.
For instance the following condition, with well over
100 nodes in its syntax tree, is the main part
of a symbolic execution query.

\begin{lstlisting}[caption={Core of an SMTLIB-formatted query arising in symbolic execution. The expression has been
    manually formatted, with numeric constants condensed from 256 bits to 8 or 16.}]
(assert (= #x01
 (bvand (my_bvlt #x01f4 SLOAD_0x13)
  (bvand (my_bvlt (bvadd #x00 #x01) var_0x1308)
   (bvand (isZero (my_eq (bvadd (bvadd #x00 #x01) #x01) (bvnot #x00)))
    (bvand (isZero (my_bvgt (bvsub #x01f4 SLOAD_0x13)
                     (bvadd (bvsub SLOAD_0x12 SLOAD_0x13) 
                            (bvnot #x00))))
     (bvand (my_bvlt (bvadd #x00 #x01) var_0x1308)
      (bvand (isZero (my_eq #x00 (bvnot #x00)))
       (bvand (isZero (my_bvgt (bvsub #x01f4 SLOAD_0x13)
                       (bvadd (bvsub SLOAD_0x12 SLOAD_0x13) 
                              (bvnot #x00))))
        (bvand (my_bvlt (bvadd #x00 #x01) var_0x1308)
         (bvand (isZero (my_eq #x00 (bvnot #x00)))
          (bvand (isZero (my_bvgt (bvsub #x01f4 SLOAD_0x13)
                          (bvadd (bvsub SLOAD_0x12 SLOAD_0x13) 
                                 (bvnot #x00))))
           (bvand (my_bvlt #x00 var_0x1308)
            (bvand (isZero (my_bvgt var_0x1308 SLOAD_0x15))
             (bvand (isZero (isZero (isZero (my_eq SLOAD_0x11 #x02))))
              (bvand (isZero 
                        (bvsub 
                          (bvand 
                            var_0x12fc
                            #x0ffffffffffffffffffffffffffffffffffffffff)
                          var_0x12fc))
               (bvand
                (isZero (my_bvslt (bvadd CALLDATASIZE0x12f3 (bvnot #x03)) 
                                  #x40))
                (isZero CALLVALUE0x12e9))))))))))))))))))
\end{lstlisting}

The free variables in this query are \sv{SLOAD\_0x11},
\sv{SLOAD\_0x12}, \sv{SLOAD\_0x13}, \sv{SLOAD\_0x15} (the values
read from the storage location at the constant address reflected in
the variable's name), \sv{var\_0x12fc}, \sv{var\_0x1308} (public function arguments),
\sv{CALLDATASIZE0x12f3} (the size in bytes of all data passed into the
function), \sv{CALLVALUE0x12e9} (the value in native cryptocurrency passed into the contract).

It is entirely straightforward for a top-down SMT solver to find a satisfying assignment
for this query. For the native solver, however, the query would only be solvable if each \sv{bvand}
constraint was handled independently, yet many constraints in the example concern
the same variable (e.g., \sv{SLOAD\_0x13}). If, instead, the query is considered as a whole,
it is too large for the native symbolic solver's bottom-up reasoning. The native solver can theoretically
handle expressions with 20 nodes, in tree form, but
only if these keep simplifying to equivalent shorter expressions. (In practice the limit
is even lower.)




\section{Related Work}\label{sec:related}

\myparagraph{\textbf{Datalog and Declarative Program Analysis.}}
Datalog-based declarative program analysis has been flourishing for at least two decades, with systems such
as bddbddb~\cite{aplas/WhaleyACL05}, Doop~\cite{oopsla/BravenboerS09}, and
Chord~\cite{chord}; and platforms such as LogicBlox~\cite{Aref:2015:DIL:2723372.2742796}, \souffle{}~\cite{Jordan16},
Inca~\cite{DBLP:conf/pldi/SzaboEB21}, and Flix~\cite{DBLP:conf/pldi/MadsenYL16}. The declarative approach has
repeatedly demonstrated that an analysis can be both scalable and easy to read. Despite starting as a database
query language, Datalog has found numerous program analysis applications in diverse domains ranging from
points-to analysis~\cite{oopsla/BravenboerS09, pldi/WhaleyL04}, to
disassemblers/decompilers~\cite{floresmontoya2019datalog,gigahorse,elipmoc}, security~\cite{Tsankov:2018:SPS:3243734.3243780} and program repair~\cite{symlog}.

\myparagraph{\textbf{Symbolic Execution and SMT Solvers.}} Symbolic execution was first proposed in
1975~\cite{mattmight:King:1976:Symbolic, boyer1975select, howden1975methodology}, but the first
scalable symbolic executors only started appearing around 2005 with DART~\cite{godefroid2005dart}
and EXE~\cite{cadar2006exe}. The timing is not a coincidence, symbolic execution relies on solving
complex symbolic constraints and the first industrial-strength SMT solvers started appearing in 2002
with CVC~\cite{stump2002cvc}, STP~\cite{ganesh2007decision} and Z3~\cite{demoura2008z3} leading the way.
SMT solver innovation led to an explosion in the symbolic execution
field with numerous new symbolic executors being proposed across diverse fields including
automatic test case generation~\cite{cadar2008klee, godefroid2012sage}, verification~\cite{jaffar2012tracer},
security~\cite{cha2012unleashing} and others~\cite{baldoni2018survey, symlog}.

\myparagraph{\textbf{Logic Programming and Solvers.}} By integrating mature SMT solvers seamlessly within Datalog evaluation,
one can hope to spur a similar wave of program analyses that will marry the two trends.
Formulog~\cite{DBLP:journals/pacmpl/Bembenek0C20} exemplifies this push, extending Datalog with functional
programming constructs to enable SMT-based program analyses, like refinement type checking and symbolic
execution. Recent Formulog work, using eager evaluation, was able to improve performance by an order of
magnitude~\cite{formulog24} for SMT-based analyses. Yet the underlying performance burdens remain: to
achieve high-performance declarative analysis, one needs to express the analysis \emph{idiomatically} in
Datalog and fully leverage a modern engine like \souffle{}. Our experiments show how large the
gap still is.


There is a wealth of work
in the field of combining constraint solving capabilities with logic-based languages. The entire sub-area
of constraint logic programming~\cite{JAFFAR1994503} extends conventional logic programming with
search over a (super-)exponential domain. In constraint logic programming, the evaluation engine
of the logic program is \emph{itself} much like an SMT solver, requiring no external integrations.
From a mile-high viewpoint, this approach has similarity with \oursystem{}. In practice,
however, constraint logic programming is not used as the basis for program analysis algorithms (e.g.,
pointer analysis).
The starting point of our work is a predictable, efficient, typically polynomial-time
Datalog evaluator. \oursystem{} enables the analysis writer to appeal to the SMT solver sparingly
and in a highly-controlled way, enabling state-of-the-art program analyses.

From the opposite end, the Z3 SMT solver contains
a fixpoint component~\cite{10.1007/978-3-642-22110-1_36} permitting it to express the kinds of analysis
algorithms one would write in Datalog. However, Datalog-like computations inside Z3 are several orders
of magnitude slower~\cite[Sec.3]{Jordan16} than a cutting-edge Datalog engine, such as \souffle{}.

Calypso~\cite{aiken2007overview} extended a Datalog variant with a packaged solver interface to
enable solving SAT and integer constraint solving in path-sensitive analyses. While Calypso was
not connected to an SMT solver at the time, the idea of natively integrating constraint solving
within the language served as an inspiration for \oursystem{}. Constructing formulas in Calypso
required the user to incrementally query external solver
predicates---e.g., \texttt{\#and} for conjunction---before appealing to the solver. Unlike Calypso,
\oursystem{} allows the user to construct formulas using native Datalog types, relies on SMTLIB for
the common constraint language, and provides custom operator support for user-defined constructs.

\myparagraph{\textbf{Solver-aided Languages.}}
Scala$^{Z3}$~\cite{koksal2011scala} expresses Z3 formula trees as native Scala
terms, utilizes the type system for capturing ill-typed constraints and allows combining Scala
evaluation with SMT solving. Scala$^{Z3}$ also provides a set of custom operators such as
\texttt{choose} and \texttt{find} to enable the user to express complex symbolic reasoning.
Rosette~\cite{torlak2013growing} is a language and system that extends Racket with
constructs that enabling symbolic reasoning and synthesis. A symbolic virtual machine
executes Rosette programs and compiles them to logical constraints that are sent to the
solver to reason about program behaviors. \oursystem{} relates to solver-aided languages in
that it also provides SMT solving capabilities for programs written in Datalog. However,
\oursystem{}'s top goal is to offer flexible solving constructs for program analysis writers
and as we witnessed in Section~\ref{sec:experiments}, SMT-only solutions may not suffice.

\myparagraph{\textbf{Analysis Platforms.}}
The best-known mixed-mode analysis platform is perhaps CPAchecker~\cite{TR002-SFU09}.
CPAchecker enables analysis of C programs using a wide range of techniques,
including abstract domains, symbolic execution, and model checking.
In addition to giving users the ability to combine these techniques,
it also provides a unified interface for analysis developers to implement their own algorithms.
This makes it easy to test and evaluate different approaches on the same platform,
enabling constructive comparison between analyses. Despite the general model, the tool
is oriented towards execution-based analyses (e.g., symbolic execution and model checking)
and not high-performance static program analyses (e.g., points-to analysis), such as those typically
implemented in Datalog.

\section{Conclusions}
\label{sec:conclusion}

We presented \oursystem, a platform for seamless integration of value-flow and symbolic
reasoning in Datalog-style declarative program analysis. Built on top of \souffle{},
\oursystem{} introduces two new capabilities for analysis writers: (1) solving complex
symbolic tasks through a set of transparent and fully customizable functors
that interact with state-of-the-art SMT solvers, and (2) solving lightweight symbolic
reasoning constraints natively through a pure Datalog solver with predictable performance. The
two complementary capabilities offer a flexible and efficient platform that enables
users to balance precision and performance by selecting the most appropriate reasoning
approach for their analysis. \oursystem{} comes with a number of useful features for
typical program analysis tasks, such as query caching, let-bindings, and support for
custom operators like \forAllStar{}.

We demonstrated the effectiveness of \oursystem{} through three case studies: a points-to
analysis, a symbolic value-flow analysis, and a complete symbolic execution engine.
Our experiments showed significant speedups compared to previously available
solutions and illustrated the practical benefits of combining lightweight and heavyweight
symbolic solvers within the same analysis pipeline. We believe that \oursystem{} is a
strong candidate for both research and industrial program analysis applications and have
open-sourced the platform to encourage further research in the field.



% ...

\bibliographystyle{plain}
\bibliography{bibliography,references,tools,bib/ptranalysis,bib/proceedings,testing-cc}

\clearpage
\appendix
\renewcommand{\thefigure}{\thesection.\arabic{figure}}
\renewcommand{\thetable}{\thesection.\arabic{table}}

\section{experimental Results}
\label{appendix:experimental-results}
\setcounter{figure}{0}

\subsection{Points-To Analysis}
\label{appendix:points-to-experiment}

\begin{figure*}[!htbp]
  \centering
  \begin{subfigure}[b]{0.8\textwidth}
      \centering
      \includegraphics[width=\textwidth]{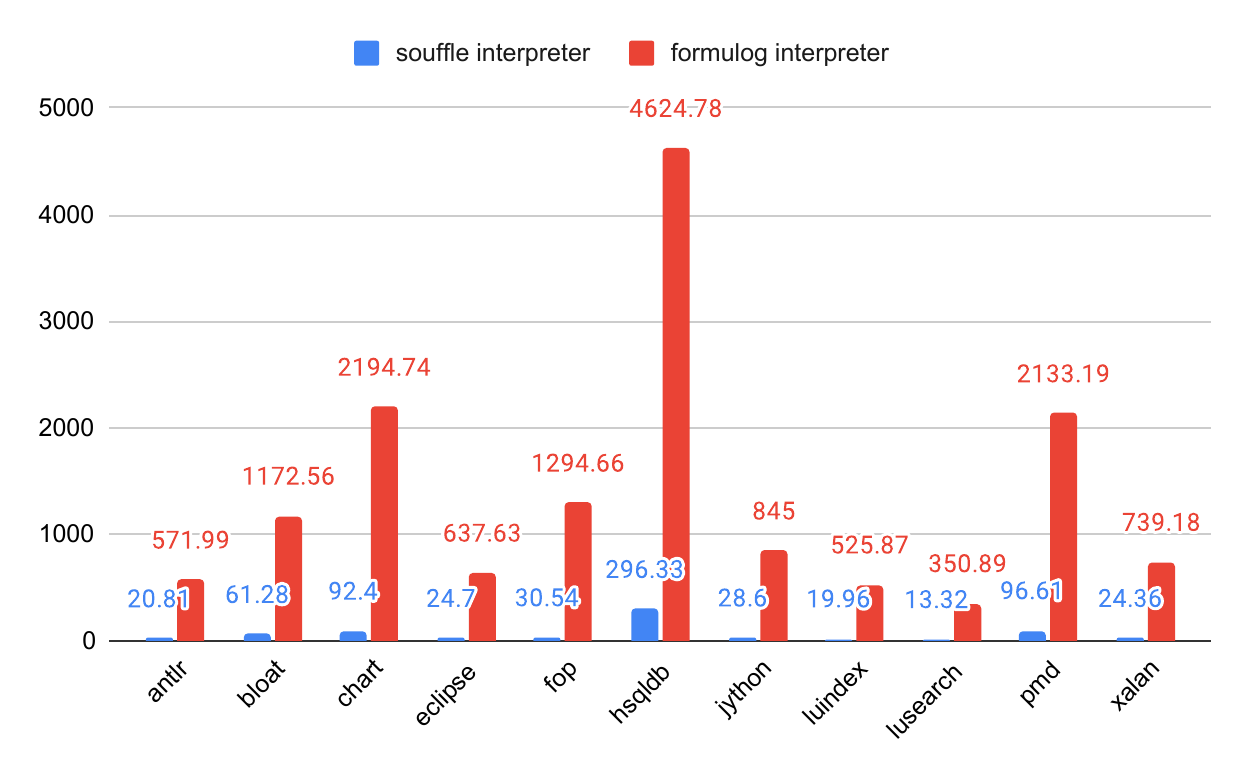}
  \end{subfigure}
  \caption[]
          {Execution time of Point-To analysis (in sec), comparison of \souffle{} vs. Formulog.
            Interpreted sequential version}
  \label{fig:times-per-benchmark-per-config-seq-interpreted}
\end{figure*}

\begin{figure*}
  \centering
  \begin{subfigure}[b]{0.8\textwidth}
      \centering
      \includegraphics[width=\textwidth]{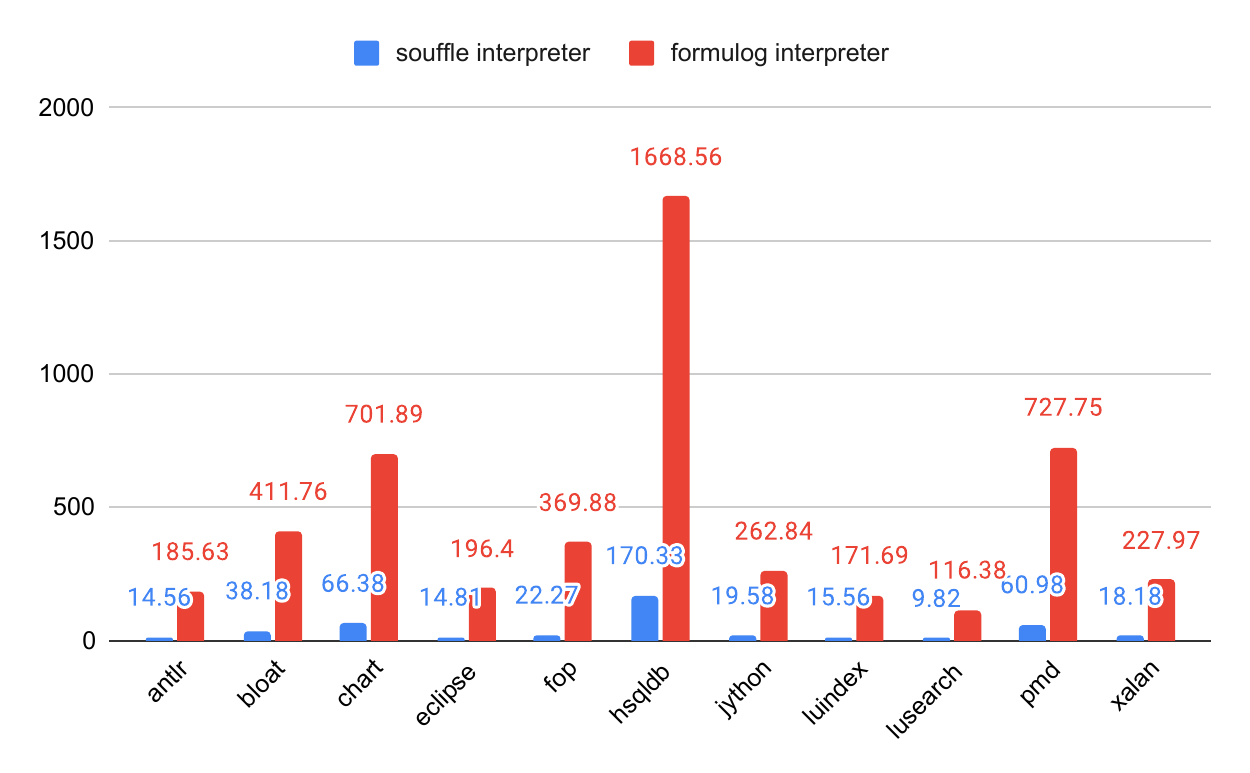}
  \end{subfigure}
  \hfill
  \vskip\baselineskip
  \begin{subfigure}[b]{0.8\textwidth}
      \centering
      \includegraphics[width=\textwidth]{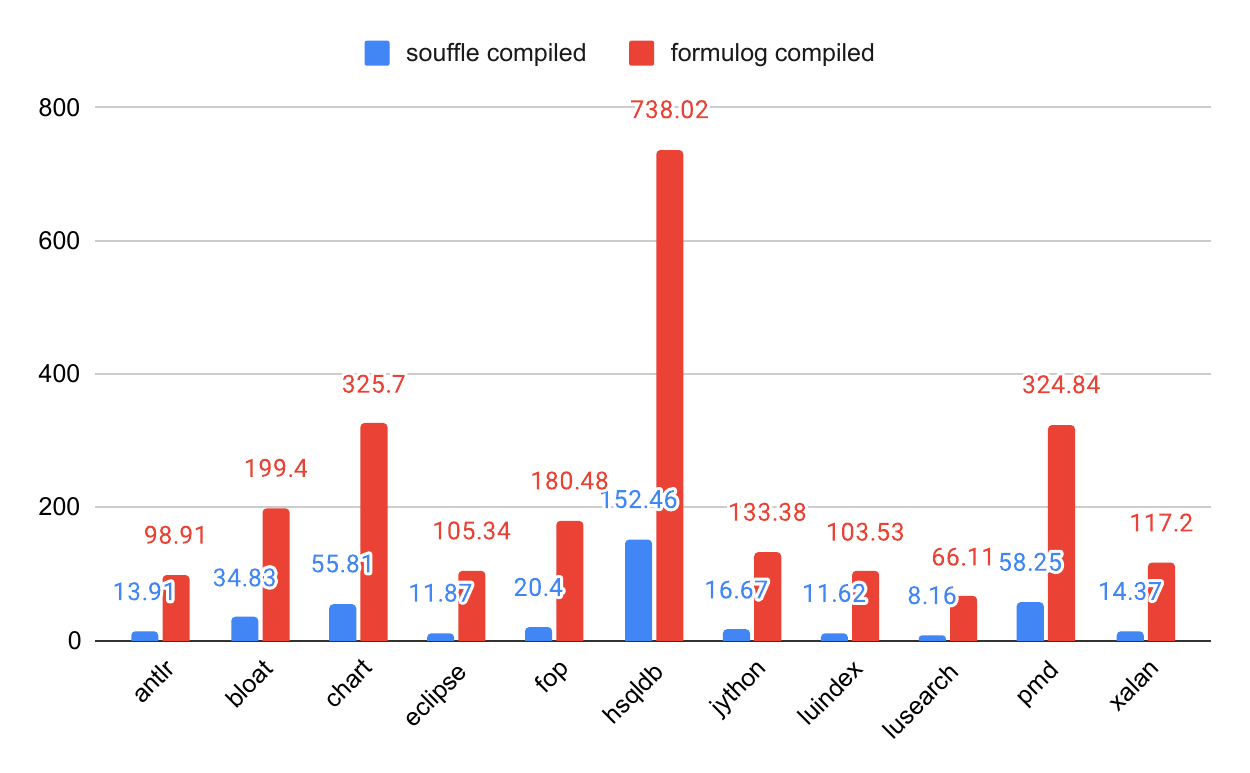}
  \end{subfigure}
  \caption[]
          {Execution time of Point-To analysis (in sec), comparison of \souffle{} vs. Formulog.
            From top to bottom: interpreted 4-threads, compiled 4-threads.}
  \label{fig:times-per-benchmark-per-config-parallel-4}
\end{figure*}

Figures~\ref{fig:times-per-benchmark-per-config-seq-interpreted} and ~\ref{fig:times-per-benchmark-per-config-parallel-4} show the average analysis times (for a single run, but we have neither
observed large variation, nor are the qualitative results affected by small changes), and complement Section~\ref{sec:experiments-points-to} by providing a more complete overview of the timing behavior.

Figures~\ref{fig:speedup-per-benchmark} and \ref{fig:average-speedup} give, in table form, the speedup per benchmark+configuration, as well as the average
speedup, respectively.

Our results indicate that, while our system outperforms Formulog in both single-threaded and 4-thread configurations,
Formulog demonstrates noticeable concurrency scaling across these configurations, reflecting an effective parallel evaluation design.

\newcommand*\rot{\rotatebox{45}}

\begin{figure}
\begin{center}
  \begin{tabular}{l|SSSS}
   \hline
 Benchmark & {Interpreted} & {Compiled} & {Interpreted} & {Compiled} \\
           & {1 thread}    & {1 thread} & {4 threads}   & {4 threads} \\
   \hline
   antlr   & 27.5 & 15.5 & 12.7 & 7.1 \\
   bloat   & 19.1 & 10.4 & 10.8 & 5.7 \\
   chart   & 23.8 & 11.9 & 10.6 & 5.8 \\
   eclipse & 25.8 & 16.8 & 13.3 & 8.9 \\
   fop     & 42.4 & 21.9 & 16.6 & 8.8 \\
   hsqldb  & 15.6 & 7.5  & 9.8  & 4.8 \\
   jython  & 29.5 & 18.2 & 13.4 & 8.0 \\
   luindex & 26.3 & 16.8 & 11.0 & 8.9 \\
   lusearch& 26.3 & 18.4 & 11.9 & 8.1 \\
   pmd     & 22.1 & 11.9 & 11.9 & 5.6 \\
   xalan   & 30.3 & 18.2 & 12.5 & 8.2 \\
   
    \hline
  \end{tabular}
\end{center}
  \caption{\oursystem{} speedup over Formulog broken down by benchmark and configuration.}
  \label{fig:speedup-per-benchmark}
\end{figure}

\begin{figure}
  \begin{center}
  \begin{tabular}{rl|S}
    \hline
    &  & {Avg. Speedup} \\
    \hline
    Interpreted & (1 Thread)  & 26.26  \\
    Compiled & (1 Thread)  & 15.25  \\
    Interpreted & (4 Threads)  & 12.23  \\
    Compiled & (4 Threads) & 7.27  \\
    \hline
  \end{tabular}
  \caption[]{Average speedup offered by \oursystem{} for each configuration.}
  \label{fig:average-speedup}
  \end{center}
\end{figure}
\end{document}

%% file: solidityenv.tex
\newcommand{\SOLIDITYKEYWORDS}

\lstnewenvironment{solidity}
    {\lstset{%
      name=main
      language=Java,         
      xrightmargin=2pt,
      xleftmargin=2pt,
      columns=fullflexible,
      keepspaces=true,
      style=numbers,
      classoffset=0,
      frame=bt,
      deletekeywords={byte, char, int, float, double},
      keywordstyle=\color{blue}\bfseries,      
      classoffset=1,
      morekeywords={modifier, mapping, contract, require, uint, emit,
                    assert, if, else, contract, function, returns, var, public}, 
      keywordstyle=\color{brown}\bfseries,  
      classoffset=0,
      commentstyle=\color{blue},             
      morecomment=[l][\color{mauve}]{//},
      morecomment=[l][\color{red}]{//!},
      morecomment=[s][\color{mauve}]{/*}{*/},
      morecomment=[s][\color{red}]{/**}{*/},
      stringstyle=\color{dkgreen},             
      escapeinside={\%*}{*)}                   
}}{}

\lstnewenvironment{solidity-small}
    {\lstset{%
      name=main
      language=Java,         
      columns=fullflexible,
      keepspaces=true,
      style=tinynonumbers,
      classoffset=0,
      frame=bt,
      deletekeywords={byte, char, int, float, double},
      keywordstyle=\color{blue}\bfseries,      
      classoffset=1,
      morekeywords={modifier, mapping, contract, require, uint, emit,
                    assert, if, else, contract, function, returns, var, public}, 
      keywordstyle=\color{brown}\bfseries,  
      classoffset=0,
      commentstyle=\color{blue},             
      morecomment=[l][\color{mauve}]{//},
      morecomment=[l][\color{red}]{//!},
      morecomment=[s][\color{mauve}]{/*}{*/},
      morecomment=[s][\color{red}]{/**}{*/},
      stringstyle=\color{dkgreen},             
      escapeinside={\%*}{*)}                   
}}{}